\documentclass[aps, prb, preprint, showpacs, amssymb, amsmath, superscriptaddress]{revtex4-1}

\usepackage{mathrsfs}
\usepackage{bm}
\usepackage{times}
\usepackage{graphicx}
\usepackage[colorlinks,citecolor=blue,linkcolor=blue]{hyperref}
\usepackage{array}
\usepackage{float}
\usepackage{colortbl}
\usepackage{multirow}
\usepackage{tabularx}

\usepackage{graphicx}% Include figure files
\usepackage{dcolumn}% Align table columns on decimal point
\usepackage{bm}% bold math
\begin{document}

\title{Large magnetic anisotropy energy and robust half-metallic ferromagnetism in 2D MnXSe$_4$ (X = As, Sb)}% Force line breaks with \\

\author{Tengfei Hu}
\affiliation{State Key Laboratory of Metastable Materials Science and Technology \& Key Laboratory for Microstructural Material Physics of Hebei Province, School of Science, Yanshan University, Qinhuangdao, 066004, China}
\author{Wenhui Wan}
\affiliation{State Key Laboratory of Metastable Materials Science and Technology \& Key Laboratory for Microstructural Material Physics of Hebei Province, School of Science, Yanshan University, Qinhuangdao, 066004, China}
\author{Yingmei Li}
\affiliation{State Key Laboratory of Metastable Materials Science and Technology \& Key Laboratory for Microstructural Material Physics of Hebei Province, School of Science, Yanshan University, Qinhuangdao, 066004, China}
\author{Yanfeng Ge}
\affiliation{State Key Laboratory of Metastable Materials Science and Technology \& Key Laboratory for Microstructural Material Physics of Hebei Province, School of Science, Yanshan University, Qinhuangdao, 066004, China}
\author{Kaicheng Zhang}
\affiliation{Department of Physics, Bohai University, Jinzhou 121000, China}
\author{Yong Liu}
\email{yongliu@ysu.edu.cn, ycliu@ysu.edu.cn}\affiliation{State Key Laboratory of Metastable Materials Science and Technology \& Key Laboratory for Microstructural Material Physics of Hebei Province, School of Science, Yanshan University, Qinhuangdao, 066004, China}

\date{\today}% It is always \today, today,

\begin{abstract}
In recent years, intrinsic two-dimensional (2D) magnetism aroused great interest because of its potential application in spintronic devices. However, low Curie temperature (\emph{T}$_c$) and magnetic anisotropy energy (MAE) limit its application prospects. Here, using first-principles calculations based on density-functional theory (DFT), we predicted a series of stable MnXSe$_4$ (X=As, Sb) single-layer. The MAE of single-layer MnAsSe$_4$ and MnSbSe$_4$ was 648.76 and 808.95 ${\mu}$eV per Mn atom, respectively. Monte Carlo (MC) simulations suggested the \emph{T}$_c$ of single-layer MnAsSe$_4$ and MnSbSe$_4$ was 174 and 250 K, respectively. The energy band calculation with hybrid Heyd-Scuseria-Ernzerhof (HSE06) function indicated the MnXSe$_4$ (X = As, Sb) were ferromagnetic (FM) half-metallic. Also it had 100\% spin-polarization ratio at the Fermi level. For MnAsSe$_4$ and MnSbSe$_4$, the spin-gap were 1.59 and 1.48 eV, respectively. These excellent magnetic properties render MnXSe$_4$ (X = As, Sb) promising candidate materials for 2D spintronic applications.

\end{abstract}

\maketitle

%\tableofcontents

\section{INTRODUCTION}

Spintronics, which uses the spin degree freedom of electrons, is significant for future information technologies owing to its great promise in enhancing data processing speed and integration densities~\cite{1,2}. Half-metals can generate 100\% spin polarized currents without any external operation and play significant roles in the spintronic applications~\cite{3,4}. Half-metallic materials with high \emph{T}$_c$ and large MAE are fundamental to build practical spintronic devices that can work at room temperature~\cite{5,6}. In order to realize this concept at the nanoscale, the development of low-dimensional half-metallic materials with the characteristics above is a key problem. Therefore, there has been a flurry of research into magnetic half-metals, such as RbSe, CsTe, TiCl$_3$, VCl$_3$, MnX (X=P, As) and so on~\cite{7,8,9,10,11,12}. However, these demonstrated half-metals have serious shortcomings, such as low \emph{T}$_c$ or MAE. Until now, intrinsic half-metallic materials with both high \emph{T}$_c$ and large MAE are still absent in experiments.

Unlike in bulk magnetic materials, the long-range magnetic ordering in 2D structures is impossible without magnetic anisotropy, which is required for counteracting thermal fluctuations~\cite{13}. MAE is defined as the difference between the energy corresponding to the magnetization in the in-plane and off-plane directions (MAE = E$_{//}$ - E$_{\perp}$). Therefore, a positive (negative) value of MAE indicates that the off-plane (in-plane) is easy axis. However, MAE mainly comes from the influence of spin-orbit coupling (SOC)~\cite{14}. It is better for the magnetic ordering to resist the heat fluctuation with larger values of MAE. Noncollinear calculation showed that single-layer CrPS$_4$ exhibited the MAE of 40.0 ${\mu}$eV per Cr atom with the spins favorably aligned along the off-plane direction~\cite{15}. Generally, long-range ferromagnetic order mainly exists in 3d transition metals and their compounds. However, 3d elements have relatively weak SOC and strong SOC can only be found in heavy elements. So we replace P with As/Sb atoms, S with Se atoms, respectively. By stability calculation, we replace Cr with Mn atoms.

In this paper, we predicted two emerging class of 2D intrinsic FM half-metallic materials (single-layer MnAsSe$_4$ and MnSbSe$_4$). Our calculations indicated that single-layer MnXSe$_4$ (X = As, Sb) crystals were mechanically and dynamically stable, so they may be synthesized experimentally. We also demonstrated that single-layer MnXSe$_4$ (X = As, Sb) had giant MAE. The single-layer MnAsSe$_4$ (MnSbSe$_4$) have abundant states at the Fermi level in one spin direction and a band gap of 1.59 eV (1.48 eV) in the opposite spin direction, and its half-metallic band gap is 0.61 eV (0.59 eV). Furthermore, we demonstrated that the MnAsSe$_4$ and MnSbSe$_4$ exhibited high \emph{T}$_c$ about 174 K and 250 K, respectively.

\section{METHODS}

Kohn-Sham DFT calculations were performed using the projector augmented wave method, as implemented in the plane-wave code VASP~\cite{16,17,18}. The plane-wave cutoff energy was set to 500 eV. A Monkhorst-Pack special k-point mesh~\cite{19} of $7\times7\times1$ for the Brillouin zone integration was found to be sufficient to obtain the convergence. We used a Perdew-Burke-Ernzerhof (PBE) type generalized gradient approximation (GGA) in the exchange-correlation functional~\cite{20}. For the MAE calculation, the SOC was included. A conjugate-gradient algorithm was employed for geometry optimization using convergence criteria of 10$^{-7}$ eV for the total energy and 0.005 eV/{\AA} for Hellmann-Feynman force components. We displayed that the results with Hubbard U term 5 eV for Mn~\cite{21} as suggested by Dudarev \emph{et al}~\cite{22}. However, for a more accurate estimation of the band structure, the hybrid functional of HSE06~\cite{23} was used. Phonon dispersions were calculated by density functional perturbation theory~\cite{24} by the Phonopy package interfaced to VASP code with a $2\times2\times1$ supercell. Moreover, a 15 {\AA} vacuum was applied along the z axis to avoid any artificial interactions between images.

\section{RESULTS AND DISCUSSION}
\begin{figure}[htb]
  \centering
  \includegraphics[width=0.5\textwidth]{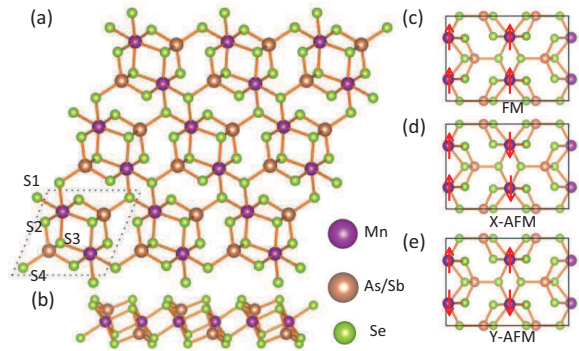}
 \caption{(Color online) (a) Top view and (b) side view of single-layer MnXSe$_4$ (X = As, Sb). The primitive cell is indicated by dotted line in (a). (c-e) Three different magnetic configurations.}\label{fig1}
\end{figure}
Structural models of single-layer MnXSe$_4$ (X = As, Sb) are shown in Fig.~\ref{fig1} (a) and (b). The primitive cell is indicated by dotted line in (a), and it contains two Mn atoms. The optimized lattice parameters for MnAsSe$_4$ (MnSbSe$_4$)
are a = b = 7.20{\AA} (7.43{\AA}). In order to identify their preferred magnetic ground-state, a FM and two antiferromagnetic (AFM) magnetic configurations were constructed. FM, X-AFM and Y-AFM are shown in Fig.~\ref{fig1} (c), (d) and (e), respectively. The energy difference ($\Delta$E) relative to FM configurations of MnAsSe$_4$ (MnSbSe$_4$) is 369.13 (263.90) and 518.83 (384.06) meV for X-AFM and Y-AFM orders, respectively. Therefore, FM is the ground-state. We calculated that each primitive cell was an integer magnetic moment of 8 $\mu_B$, and the local magnetic moment per Mn atom was about 4 $\mu_B$. The magnetic moment is consistent with the +3 oxidation state of Mn, so it is 4s$^0$3d$^4$ electronic configuration. The unpaired d electrons contribute the magnetism. According to the Hund's rule and the Pauli Exclusion Principle, the four unpaired d electrons left result in the magnetic moment of 4 $\mu_B$ per Mn atom.

%The best place to locate the table environment is directly after its first reference in text
\begin{table}[h]%The best place to locate the table environment is directly after its first reference in text
 \newcommand{\tabincell}[2]{\begin{tabular}{@{}#1@{}}#2\end{tabular}}
    \centering
\begin{ruledtabular}
\caption{Bond lengths ({\AA}) of single-layer MnXSe$_4$ (X = As, Sb).}\label{table1}
\begin{tabular}{lcccccccc}
   & d$_{Mn-S1}$ & d$_{Mn-S2}$ & d$_{Mn-S3}$ & d$_{X-S2}$ & d$_{X-S3}$ & d$_{X-S4}$   \\
   \hline
   MnAsSe$_4$   & 2.76  &2.61  &2.77 &2.34  &2.39 & 2.36  \\
   MnSbSe$_4$   & 2.77  &2.62  &2.79 &2.51  &2.56 & 2.53  \\
\end{tabular}
\end{ruledtabular}
\end{table}

\begin{table}[h]%The best place to locate the table environment is directly after its first reference in text
 \newcommand{\tabincell}[2]{\begin{tabular}{@{}#1@{}}#2\end{tabular}}
    \centering
\begin{ruledtabular}
\caption{Elastic constants (N/m), Young¡¯s modulus (N/m), MAE (${\mu}$eV/Mn) and Easy axis for single-layer MnXSe$_4$ (X = As, Sb).}\label{table2}
\begin{tabular}{lcccccc}
   & C$_{11}$ & C$_{12}$ & C$_{22}$   & ${Y_{2D}}$  &MAE & Easy axis\\
   \hline
   MnAsSe$_4$ & 64.41 & 11.81 & 57.09 & 62.24  &648.76 & c\\
   MnSbSe$_4$ & 62.63 & 12.18 & 59.41 & 60.26  &808.95 & c\\
\end{tabular}
\end{ruledtabular}
\end{table}
The structural parameters of MnXSe$_4$ (X = As, Sb) are summarized in Table~\ref{table1}. The bond lengths increase from MnAsSe$_4$ to MnSbSe$_4$, it makes sense because the atomic radius of Sb is larger than the As.

The MAE listed in Table~\ref{table2}, it is important to determine the thermal stability of magnetic ordering. The MAE of MnAsSe$_4$ and MnSbSe$_4$ is 648.76 and 808.95 ${\mu}$eV/Mn, which is larger than the previous research on CrPS$_4$ (40 ${\mu}$eV/Cr)~\cite{15}. Clearly, the easy axis of these crystals are along the z direction.
Next, we determined their mechanical stability by calculating the three independent elastic constants. As shown in Table~\ref{table2}, the C$_{11}$ is 64.41 (62.63) N/m, C$_{12}$ is 11.81 (12.18) N/m, and C$_{22}$ is 57.09 (59.41) N/m for MnAsSe$_4$ (MnSbSe$_4$), respectively. Compared to CrPS$_4$, MnXSe$_4$ (X = As, Sb) exhibits a similar C$_{11}$, a smaller C$_{12}$ and a larger C$_{22}$~\cite{25}. The elastic constants fulfill the Born criteria of stability~\cite{26}, i.e., C$_{11}{>}$0, C$_{22}{>}$0 and C$_{11}$-C$_{12}{>}$0, indicating that they are mechanically stable. In-plane stiffness can be calculated by using the relation, ${Y_{2D}} = \frac{{(C_{11}^2 - C_{12}^2)}}{{{C_{11}}}}$, ${Y_{2D}}$ decreases from MnAsSe$_4$ to MnSbSe$_4$. As in-plane stiffness is a measure of rigidity, the decrease of ${Y_{2D}}$ indicates softening of the crystal~\cite{27}.
\begin{figure}[htb]
\centering
\includegraphics[width=0.5\textwidth]{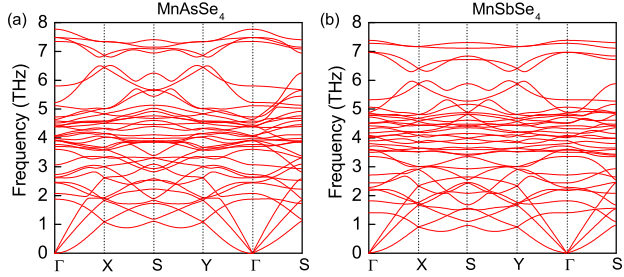}
\caption{(Color online) Theoretical phonon spectrum of single-layer MnAsSe$_4$ (a) and MnSbSe$_4$ (b) obtained from DFT calculations.}\label{fig2}
\end{figure}

To further confirm the stability of free-standing single-layer MnXSe$_4$ (X = As, Sb), we calculated their phonon dispersion. As shown in Fig.~\ref{fig2}, two materials have the same overall shape with no imaginary modes, it suggests that single-layer MnXSe$_4$ (X = As, Sb) are dynamically stable and can exist as free-standing 2D crystals.

\begin{figure}[htb]
  \centering
  \includegraphics[width=0.5\textwidth]{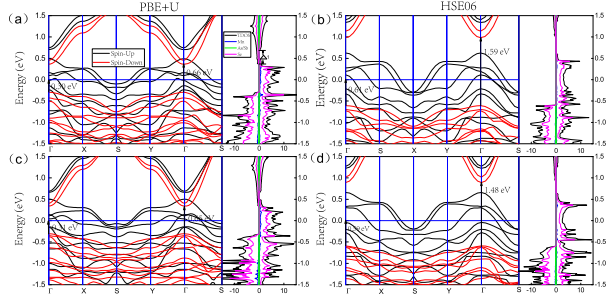}
 \caption{(Color online) Electronic band structures and DOS for single-layer MnAsSe$_4$ ((a) and (b)) and MnSbSe$_4$ ((c) and (d)). (a) and (c) using PBE+U functional, (b) and (d) using HSE06 functional.}\label{fig3}
\end{figure}

Then, the electronic properties of single-layer MnXSe$_4$ (X = As, Sb) were investigated. Since the PBE functional usually underestimates the energy band gap, a hybrid functional in the form of the HSE06 functional was used to obtain accurate electronic structures. Fig.~\ref{fig3} (a) and (c) employ the PBE+U functional, (b) and (d) employ the HSE06 functional. Notably, all the band structures show the spin-up channel crosses the Fermi level, while the spin-down channel acts as a semiconductor, indicating that they are intrinsic FM half-metallic materials with 100\% spin-polarization ratio. The most of previously discovered FM half-metals are realized by external conditions such as pressure and doping, the number of intrinsic 2D FM half-metallic materials are small~\cite{28,29,30}. Here we found that single-layer MnXSe$_4$ (X = As, Sb) were intrinsic FM half-metals, which offer more suitable candidate materials for actual nanoscaled spintronic applications. To achieve the great prospect for half-metallic in spintronic devices, wide half-metallic gap is extremely important~\cite{31,32}. Herein, the PBE+U values of the half-metallic gap are 0.30 eV and 0.31 eV for MnAsSe$_4$ and MnSbSe$_4$, respectively. For HSE06 functional, the half-metallic band gap are 0.61 and 0.59 eV for MnAsSe$_4$ and MnSbSe$_4$, respectively, which is large enough to efficiently prevent the thermally agitated spin-flip transition. The HSE06 value of the spin-gap is 1.59 (1.48) eV for the MnAsSe$_4$ (MnSbSe$_4$), which is wide enough to prevent spin leakage.

As shown in Fig.~\ref{fig3}, the densities of states (DOS) of MnXSe$_4$ (X = As, Sb) compounds are generally similar in shape. The contribution of MnXSe$_4$ (X = As, Sb) spin-down DOS mainly comes from Se atoms, while the contribution of other atoms is little. Larger spin exchange splitting is crucial for the application in spin-polarized carrier injection and detection. For the single-layer MnAsSe$_4$ and MnSbSe$_4$, a large spin exchange splitting of 0.30 and 0.31 eV in the conduction band are observed (labeled as $\triangle$1 in Fig.~\ref{fig3} (a)). The HSE06 values of the spin exchange splitting are 0.36 and 0.33 eV for the single-layer MnAsSe$_4$ and MnSbSe$_4$, respectively, which are larger than the CrGeTe$_3$ (0.24 eV)~\cite{33}. Further experiments should be conducted to clarify these interesting electronic character of MnXSe$_4$ (X = As, Sb).

\begin{figure}[htb]
    \centering
  \includegraphics[width=0.5\textwidth]{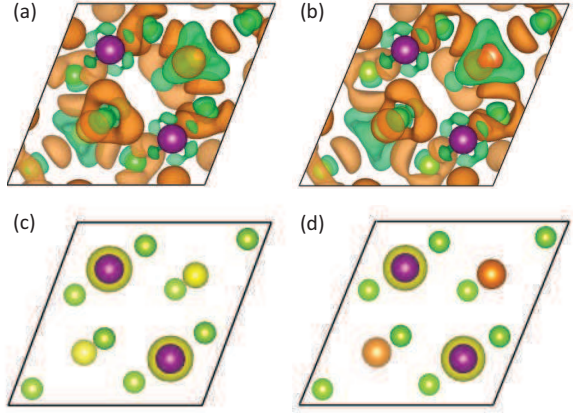}
    \caption{(Color online) (a) and (b), Charge density difference of single-layer MnAsSe$_4$ and MnSbSe$_4$. The brown (green) region represents the net charge gain (loss). The isosurface value is 0.006 e\AA$^{-3}$. MnAsSe$_4$ (c) and MnSbSe$_4$ (d), Isosurface of spin density with an isovalue of 0.05 e\AA$^{-3}$.}\label{fig4}
  \end{figure}

Fig.~\ref{fig4} (a) and (b) show the charge density difference of single-layer MnAsSe$_4$ and MnSbSe$_4$. It is defined as the difference between the charge density at the bonding point and the atomic charge density at the corresponding point. The brown (green) region represent the charge accumulation (depletion). Significant charge redistributions were observed for Mn, As/Sb and Se atoms, where the Mn and As/Sb atoms lost electrons, while the Se atoms gained electrons, meaning that there allows Mn-S bonding to be more ionic. This redistribution makes sense because Se atoms are more electronegative. The spin densities for MnAsSe$_4$ and MnSbSe$_4$ are shown in Fig.~\ref{fig4} (c) and (d), one can observe that the spin-polarization mainly comes from Mn atoms while the As/Sb and Se atoms are very small, which is consistent with the magnetic moment analysis.

\begin{figure}[htp]
  \centering
  \includegraphics[width=0.5\textwidth]{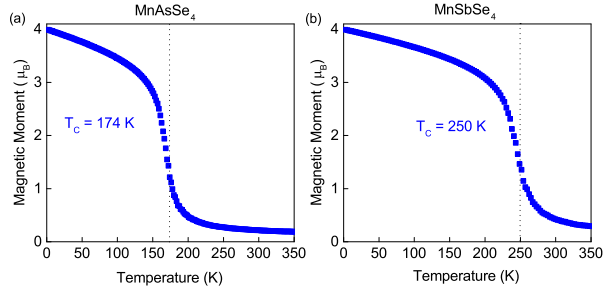}
 \caption{(Color online) On-site magnetic moment of Mn atoms versus temperature in single-layer MnAsSe$_4$ (a) and MnSbSe$_4$ (b)  based on the MC simulation.}\label{fig5}
\end{figure}

The \emph{T}$_c$ is a key parameter to the practical application of spintronic devices. Therefore, it is necessary to understand the behavior of the magnetism with temperature. Based on the Heisenberg model, the \emph{T}$_c$ of MnXSe$_4$ (X = As, Sb) single-layer can be estimated by using MC simulations:
${H =  - \sum\limits_{ < ij > } {{J_{ij}}{S_i}{S_j}}-\sum\limits_{i}A({S_{i}^{z}})^{2}}$, where $\emph{J}_{ij}$ is the exchange coupling parameter of nearest-neighbor Mn-Mn pairs, $\emph{S}_i$ represent the spin of atom $\emph{i}$, \emph{A} is anisotropy energy parameter and $S_{i}^{z}$ is the spin component along the z direction. In order to compute the $\emph{J}$, the energies of different magnetic configurations can be regarded as
\begin{equation}
{E_{X-AFM}} = {E_0} + 4{J_x}{S^2} - 4{J_y}{S^2}-\emph{A}{S^2}, \vspace{-1ex}
\end{equation}
\begin{equation}
{E_{FM}} = {E_0} - 4{J_x}{S^2} - 4{J_y}{S^2}-\emph{A}{S^2},
\end{equation}
\begin{equation}
{E_{Y-AFM}} = {E_0} - 4{J_x}{S^2} + 4{J_y}{S^2}-\emph{A}{S^2}.
\end{equation}
 The quantity \emph{E$_0$} is the ground-state energy of single-layer MnXSe$_4$ (X = As, Sb) without spin polarizations~\cite{34} and S=2. Our results show that the total $\emph{J}_{ij}$ of MnAsSe$_4$ (MnSbSe$_4$) along the x and y directions are 11.54 (8.25) meV and 16.21 (12.00) meV, respectively. The functional relationship between magnetic moment and temperature is shown in Fig.~\ref{fig5}, at absolute zero temperature, all the spin of Mn atoms point in the same direction, forming a strict FM order, while the magnetic moment decreases rapidly when heated. The critical point from FM to paramagnetic transition was around 174 K and 250 K for MnAsSe$_4$ and MnSbSe$_4$, respectively. It is significantly higher than those reported before, e.g., CrI$_3$ monolayer (45 K)~\cite{35} and MnSTe monolayer (85 K)~\cite{36}.

%\begin{figure}
%\includegraphics{1.eps} % Here is how to import EPS art
%\caption{\label{fig1} A figure caption. The figure captions are
%aut
%omatically numbered.}
%\end{figure}
\section{CONCLUSION}
In summary, we report two experimentally viable 2D intrinsic ferromagnetic half-metallic materials (MnAsSe$_4$ and MnSbSe$_4$ single-layer), which expose appreciable MAE about 648.76 and 808.95 ${\mu}$eV/Mn. The MnXSe$_4$ (X = As, Sb) are mechanically and dynamically stable. We further reveal that MnXSe$_4$ (X = As, Sb) show a high \emph{T}$_c$ about 174 K and 250 K, respectively. The band structures show that single-layer MnXSe$_4$ (X = As, Sb) have a 100\% spin-polarization ratio at the Fermi level. Also, for the semiconducting channel, the spin-gap are 1.59 eV and 1.48 eV for MnAsSe$_4$ and MnSbSe$_4$, respectively. The intrinsic half-metallic with high \emph{T}$_c$ and large MAE confer single-layer MnXSe$_4$ (X = As, Sb) as a promising functional materials for spintronic applications. We hope our study will stimulate further experimental effort in this subject.

\begin{acknowledgments}
This work was supported by National Natural Science Foundation of China (No.11904312 and 11904313), the Project of Department of Education of Hebei Province, China(No.ZD2018015 and QN2018012) and the Natural Science Foundation of Hebei Province (No. A2019203507). Thanks to the High Performance Computing Center of Yanshan University.
\end{acknowledgments}

\nocite{*}
\bibliographystyle{elsarticle-num-names}

\bibliography{20200712Maunscript}% Produces the bibliography via BibTeX.
\end{document}